\documentclass[aps,pra,twocolumn,superscriptaddress,floatfix,longbibliography]{revtex4-2}

\usepackage{graphicx}
\usepackage{amsmath}
\usepackage{amssymb}
\usepackage{bm}

\graphicspath{{figs/}}
\newcommand{\dd}{\mathrm{d}}
\begin{document}

\title{Polarization instability of unpolarized excited states in trapped two-component Bose--Einstein condensates}

\author{Zihao Liu}
\affiliation{
Hebei Key Laboratory of Physics and Energy Technology, Department of Mathematics and Physics, North China Electric Power University, Baoding, Hebei 071003, China
}
\author{Sulun Du}
\affiliation{
Hebei Key Laboratory of Physics and Energy Technology, Department of Mathematics and Physics, North China Electric Power University, Baoding, Hebei 071003, China
}
\author{Mingshu Zhao}
\email{zmshum@ncepu.edu.cn}
\affiliation{
Hebei Key Laboratory of Physics and Energy Technology, Department of Mathematics and Physics, North China Electric Power University, Baoding, Hebei 071003, China
}
\date{\today}

\begin{abstract}
A binary Bose--Einstein condensate in an elongated trap supports unpolarized excited states, in which both components share the same spatial profile.
Within mean-field theory we show that the interspecies repulsion at which these states demix rises with the excitation level.
Since each component keeps its atom number, the components separate by moving in opposite directions in the trap, and demixing sets in when the interaction energy released by the separation balances its cost in kinetic and trap energy.
We show that the threshold also marks the onset of an exponential, non-oscillating growth of the separation, which divides the cloud into domains of opposite polarization, with walls at the density nodes and further walls interlaced with them.
In the Thomas--Fermi regime of a quasi-one-dimensional condensate the rise of the threshold follows power laws in the ratio of the trap oscillator energy to the chemical potential.
The trap raises the ground-state threshold, in units of the intraspecies repulsion, by 1.82 times the square of this ratio.
Each density node of an excited state near the trap center raises the threshold further by 1.24 times the cube of this ratio.
Because the steps between successive levels scale as the cube of this ratio, they fall as the inverse square of the atom number, and the dependence on the excitation level is a mesoscopic effect, appreciable only when the chemical potential is no more than a few times the oscillator energy.
In a hyperfine mixture of $^{87}$Rb a Feshbach resonance tunes the interspecies repulsion across the thresholds of the lowest states.
\end{abstract}

\maketitle

\section{Introduction}
\label{sec:intro}

Binary Bose--Einstein condensates in elongated traps support stationary excited states in which both components occupy the same spatial profile; the simplest of them is the dark--dark soliton~\cite{Hoefer2011,Yan2012,Morera2018}.
Since the two components have equal densities at every point, these states are unpolarized, and their components separate once the interspecies repulsion is strong enough.
Dark solitons and their trains have been generated in elongated condensates~\cite{Becker2008,Theocharis2010}, and in a hyperfine mixture of $^{87}$Rb a Feshbach resonance has tuned the interspecies repulsion across the demixing transition~\cite{Tojo2010,Nicklas2011}.
For a uniform mixture, mean-field theory places the separation of the mixed ground state where the interspecies repulsion exceeds the geometric mean of the intraspecies repulsions, $g_{12}^{2}>g_{11}g_{22}$~\cite{HoShenoy1996,Timmermans1998,PuBigelow1998,AoChui1998}.
At what interspecies repulsion, then, do the unpolarized excited states of a trapped condensate demix, how does this threshold depend on the number of density nodes, and does the separation begin to grow in time at the same threshold?

Excited states of binary condensates have been studied mainly as solitary waves in which the two components occupy distinct modes, such as dark--bright solitons and their bound states~\cite{BuschAnglin2001,Becker2008,Hamner2011,Middelkamp2011,Yan2011,Katsimiga2017,Gao2026}, beating dark--dark solitons~\cite{Charalampidis2016,Yan2012}, trapped dark--dark states with different node numbers in the two components~\cite{Wang2021,Xu2017}, dark--antidark structures~\cite{Danaila2016,Katsimiga2020}, dark solitons in one component of a mixed or phase-separated binary condensate~\cite{OhbergSantos2001,RoyAngom2014}, and polarized dark solitons and domain walls in nonlinear optics and in binary condensates~\cite{HaeltermanSheppard1994,SheppardKivshar1997,CoenHaelterman2001}.
In all of these states the two components have different spatial profiles, and the states are polarized from the outset.
The dark--dark soliton with coincident notches, whose components share one profile, has been analyzed on the miscible side, where windows of oscillatory instability extend up to an interspecies repulsion close to the intraspecies one, and for an attractive interspecies coupling, where its two notches separate through a pitchfork~\cite{Morera2018}.

For the mixed ground state the trap already shifts the onset of separation away from the uniform criterion, since the separating components must pay kinetic and trap energy~\cite{EsryGreene1999,AoChui1998,Trippenbach2000,RiboliModugno2002,Lee2016,Wen2020,PyzhSchmelcher2020,Baizakov2025}, and in tight 1D confinement the kinetic energy keeps the components mixed beyond the uniform criterion~\cite{Wen2012}; in a quasi-2D trap the same quantum pressure shifts the transition and enhances the spin fluctuations on its miscible side~\cite{Bisset2018}.
The shift is controlled by the ratio of the oscillator quantum to the chemical potential, the polarized ground state emerges continuously through a supercritical pitchfork, and phase-separated states of increasing complexity branch off the mixed state at stronger couplings~\cite{Navarro2009}.
Since the mixed ground state is a minimum of the energy, its energetic threshold is also a dynamical one, at which the frequency of the spin-dipole mode, the out-of-phase oscillation of the two components, softens to zero~\cite{Ticknor2013}; this softening has been observed on the miscible side of the transition in a sodium mixture~\cite{Bienaime2016}.

In this paper we determine the interspecies repulsion $g_{12}^{*}(n)$ at which the unpolarized state with $n$ nodes demixes.
For equal intraspecies couplings $g_{11}=g_{22}\equiv g$, the interaction energy does not depend on the local composition at the SU(2)-symmetric point $g_{12}=g$, and for $g_{12}>g$ separating the components lowers it.
Each component keeps its atom number, however, and the two components therefore separate by moving in opposite directions in the trap, at a cost in kinetic and trap energy; demixing sets in when the released interaction energy balances this cost, and the threshold rises monotonically with $n$ at every coupling computed.

In the Thomas--Fermi (TF) regime, with $\mu$ the chemical potential of the mixed state, the trap correction to the ground-state threshold approaches a universal multiple of $(\hbar\omega/\mu)^{2}$, a scaling that a Gaussian variational treatment also finds~\cite{Navarro2009}.
A dark notch near the trap center removes a narrow segment from the region available to the polarization, which raises the cost, and expels atoms, which raise $\mu$ and lower it, and the net effect multiplies the correction by one plus a universal multiple of $\hbar\omega/\mu$.
The step between successive levels in $g_{12}/g$ is thus of third order in $\hbar\omega/\mu$ and falls as the inverse square of the atom number, a mesoscopic effect that is appreciable only when $\mu$ is no more than a few $\hbar\omega$, as for a hundred to a few hundred atoms per component in typical elongated traps; there the computed thresholds give the steps, and the asymptotic laws their order of magnitude.
The scaling of the ground-state correction, the softening of the spin-dipole mode that underlies it and the counting of unstable modes through the Krein signature are known~\cite{Navarro2009,Ticknor2013,KKS2004,Pelinovsky2005}; the coefficients and the dependence on $n$ are new.

Because a moving dark soliton is shallower and carries less energy, the excited states, unlike the ground state, are saddles of the energy even below the threshold~\cite{Theocharis2010,Morera2018}, and the energy balance alone does not decide whether the separation grows in time.
The linearized dynamics shows that the energetic threshold is also the onset of an exponential, non-oscillating growth of the separation, and the polarized state emerges continuously through a supercritical pitchfork; for the dark--dark soliton at the chemical potential of Ref.~\cite{Morera2018}, the spectrum continued beyond the SU(2) point is stable up to this threshold.
For the hyperfine pair $|1,+1\rangle$ and $|2,-1\rangle$ of $^{87}$Rb~\cite{Tojo2010}, a Feshbach resonance tunes the interspecies repulsion across the thresholds of the lowest states.

\section{Energy balance of demixing}
\label{sec:landau}

\subsection{Model}

We describe the mixture in an elongated trap by the quasi-1D Gross--Pitaevskii (GP) energy functional
\begin{equation}
\begin{aligned}
E=\int\Big[&\frac{\hbar^{2}}{2m}\big(|\psi_{1}'|^{2}+|\psi_{2}'|^{2}\big)+V\big(|\psi_{1}|^{2}+|\psi_{2}|^{2}\big)\\
&+\tfrac{g}{2}\big(|\psi_{1}|^{4}+|\psi_{2}|^{4}\big)+g_{12}|\psi_{1}|^{2}|\psi_{2}|^{2}\Big]\dd x,
\end{aligned}
\label{eq:energy}
\end{equation}
where $m$ is the atomic mass, $V=\tfrac{1}{2}m\omega^{2}x^{2}$ and $g_{11}=g_{22}\equiv g$.
The atom numbers of the components are conserved separately, and each component is normalized to unity, $\int|\psi_{j}|^{2}\,\dd x=1$.
The reduction to 1D assumes a transverse trap frequency $\omega_{\perp}\gg\omega$, with the transverse motion frozen in its ground state.
With $N$ atoms in each component and scattering lengths $a_{jk}$ the couplings are $g_{jk}=2N\hbar\omega_{\perp}a_{jk}$.
Numerical results are given in oscillator units, with energies in units of $\hbar\omega$, lengths in units of $a_{\rm ho}=(\hbar/m\omega)^{1/2}$ and couplings in units of $\hbar\omega\,a_{\rm ho}$, in which $g_{jk}=2Na_{jk}\omega_{\perp}/(\omega a_{\rm ho})$.
Stationary states are constrained critical points of Eq.~(\ref{eq:energy}), with the chemical potentials of the two components as Lagrange multipliers, and their wave functions can be taken real, with a change of sign at each node.

The unpolarized states have $\psi_{1}=\psi_{2}=\phi_{n}$, and both stationary equations reduce to
\begin{equation}
-\frac{\hbar^{2}}{2m}\phi_{n}''+V\phi_{n}+g_{\rm eff}\phi_{n}^{3}=\mu_{n}\phi_{n},
\qquad g_{\rm eff}=g+g_{12},
\label{eq:single}
\end{equation}
the stationary equation of a single condensate with coupling $g_{\rm eff}$.
Its $n$th excited state $\phi_{n}$, with $n$ nodes and chemical potential $\mu_{n}$, is the unpolarized state with $n$ dark notches.

\subsection{Polarization at fixed atom numbers}

The local composition of the mixture, the fraction of each component at a point, is described by the spin density $s=|\psi_{1}|^{2}-|\psi_{2}|^{2}$, the local magnetization of the mixture regarded as a pseudospin-$1/2$ condensate.
With the total density $\rho=|\psi_{1}|^{2}+|\psi_{2}|^{2}$ the interaction energy becomes
\begin{equation}
E_{\rm int}=\frac{1}{4}\int\Big[(g+g_{12})\,\rho^{2}+(g-g_{12})\,s^{2}\Big]\dd x,
\label{eq:nsdecomp}
\end{equation}
in which the composition enters through the single coefficient $g-g_{12}$; at the SU(2)-symmetric point $g_{12}=g$ the composition drops out, and for $g_{12}>g$ a separation of the components at fixed total density lowers the energy.

Let $u$ be a real profile, normalized, $\int u^{2}\,\dd x=1$, and orthogonal to the condensate, $\int\phi_{n}u\,\dd x=0$, and set
\begin{equation}
\psi_{1,2}=\frac{\phi_{n}\pm\varepsilon u}{\sqrt{1+\varepsilon^{2}}}.
\label{eq:ansatz}
\end{equation}
The orthogonality condition keeps both components normalized for every $\varepsilon$, so that the ansatz conserves $N_{1}$ and $N_{2}$.
The spin density $s=4\varepsilon\phi_{n}u/(1+\varepsilon^{2})$ then integrates to zero; where one component gains density the other loses it, and the two components move in opposite directions in the trap.

With the deformations that are symmetric in the two components, such as the change of the total density, made stationary at fixed $\varepsilon$, the energy depends on $\varepsilon$ alone and is even in it, $E(\varepsilon)=E_{0}+\alpha\varepsilon^{2}+\beta\varepsilon^{4}+\cdots$, since exchanging the components maps $\varepsilon\to-\varepsilon$ and leaves Eq.~(\ref{eq:energy}) invariant.
This relaxation is of order $\varepsilon^{2}$ and changes the energy only at order $\varepsilon^{4}$, and $\alpha$ therefore follows from the energy along the path of Eq.~(\ref{eq:ansatz}), which Appendix~\ref{app:a} evaluates,
\begin{equation}
\alpha[u]=2\big[T_{u}+V_{u}+(3g-g_{12})M-\mu_{n}\big],
\label{eq:a}
\end{equation}
with $T_{u}=(\hbar^{2}/2m)\int(u')^{2}\,\dd x$, $V_{u}=\int Vu^{2}\,\dd x$ and $M=\int\phi_{n}^{2}u^{2}\,\dd x$.

\begin{figure}[t]
\includegraphics{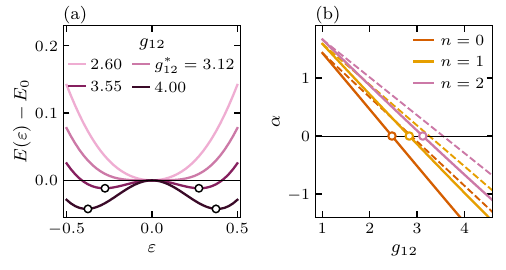}
\caption{Energy cost of polarization at $g=1$. (a) Energy along the path of Eq.~(\ref{eq:ansatz}) with $u=u_{n+1}$ for $n=2$ at four values of $g_{12}$, including the threshold $g_{12}^{*}$, in shades of the color of $n=2$ in (b); circles mark the minima of the path energy above threshold, which differ from the polarized stationary states since the path leaves the total density unrelaxed. (b) Quadratic coefficient $\alpha$ against $g_{12}$ for $n=0,1,2$. Solid lines give the full calculation, in which the state $\phi_{n}$ and the polarization mode $u_{n+1}$ are recomputed at each $g_{12}$; dashed lines give the estimate of Eq.~(\ref{eq:criterion}), in which both are held at their form at the SU(2) point, so that $\alpha$ falls linearly with $g_{12}$. The two coincide at $g_{12}=g$, and circles mark the thresholds, where the solid lines cross zero.}
\label{fig:landau}
\end{figure}

\subsection{Threshold of a saddle}
\label{sec:criterion}

The threshold is the coupling at which a polarized branch splits off from the unpolarized state, and it follows from the spectrum of the quadratic form of Eq.~(\ref{eq:a}).
Integrating the kinetic term by parts, $\alpha=2\lambda[u]$ with $\lambda[u]=\int u\hat{A}u\,\dd x$ and
\begin{equation}
\hat{A}=-\frac{\hbar^{2}}{2m}\partial_{x}^{2}+V+(3g-g_{12})\phi_{n}^{2}-\mu_{n}.
\label{eq:Ahat}
\end{equation}
Making $\lambda[u]$ stationary under the normalization and orthogonality conditions gives the constrained eigenvalue problem $\hat{A}u=\lambda u+c\,\phi_{n}$ with $\int\phi_{n}u\,\dd x=0$, derived in Appendix~\ref{app:constr}, whose eigenfunctions $u_{k}$, labeled by increasing eigenvalue with $k=0,1,2,\dots$ and the label $n$ of $\phi_{n}$ omitted, diagonalize the quadratic energy, $E=E_{0}+2\sum_{k}\lambda_{k}\varepsilon_{k}^{2}+O(\varepsilon^{4})$, with $\varepsilon_{k}$ the amplitude along $u_{k}$.

A polarized branch splits off where a constrained eigenvalue $\lambda_{k}$ passes through zero.
Along that direction the energy is $E_{0}+2\lambda_{k}\varepsilon^{2}+\beta\varepsilon^{4}$, and for $\beta>0$ the unpolarized state turns, along this direction, from a minimum into a maximum flanked by two polarized minima at $\varepsilon^{2}=-\lambda_{k}/\beta$, which grow continuously out of it; Fig.~\ref{fig:landau}(a) shows this demixing transition for $n=2$ at $g=1$.
A constrained eigenvalue that is negative at every $g_{12}$ creates no new state, and the unpolarized excited states are saddles at every $g_{12}$, since already at $g_{12}=0$, where the components do not interact, $\hat{A}$ has $n$ negative constrained eigenvalues, which reflect that $\phi_{n}$ is an excited state and are unrelated to miscibility.

With $\delta g=g_{12}-g$ the coefficient of $\phi_{n}^{2}$ in $\hat{A}$ is $3g-g_{12}=g_{\rm eff}-2\,\delta g$, and
\begin{equation}
\hat{A}=\hat{B}-2\,\delta g\,\phi_{n}^{2},
\qquad
\hat{B}=-\frac{\hbar^{2}}{2m}\partial_{x}^{2}+V+g_{\rm eff}\phi_{n}^{2}-\mu_{n},
\label{eq:split}
\end{equation}
where $\hat{B}\phi_{n}=0$ at every coupling by Eq.~(\ref{eq:single}), so that $\hat{B}$ is the single-particle Hamiltonian in the self-consistent potential $V+g_{\rm eff}\phi_{n}^{2}$, measured from $\mu_{n}$, whose $n$th orbital is $\phi_{n}$.
We denote by $\mathcal{N}_{-}(\cdot)$ the number of negative eigenvalues of an operator on the complement of $\phi_{n}$.
Below the SU(2) point, where $\delta g\leq0$, $\hat{A}$ lies above $\hat{B}$, whose count is $\mathcal{N}_{-}(\hat{B})=n$ at every $g_{12}$, $\phi_{n}$ being its zero mode with $n$ nodes, and the min--max principle gives $\mathcal{N}_{-}(\hat{A})\leq n$.
We therefore define the threshold $g_{12}^{*}$ as the smallest $g_{12}>g$ at which $\mathcal{N}_{-}(\hat{A})$ exceeds $n$; at every coupling and excitation level considered here the computed count equals $n$ at every coupling sampled between $0$ and $g_{12}^{*}$, so that no constrained eigenvalue is found to cross zero below the threshold, and $n+1$ immediately above it.
In every case computed the eigenvalue that crosses zero belongs to the mode $u_{n+1}$ with $n+1$ nodes, whose parity is opposite to that of $\phi_{n}$, as Appendix~\ref{app:constr} shows; for such modes the orthogonality constraint is inactive, and the threshold is the zero of an ordinary eigenvalue of $\hat{A}$ in that parity sector.

\subsection{Estimate at the SU(2) point}
\label{sec:su2}

At the SU(2) point the interaction energy does not depend on the composition, and the cost of a polarization is set by the single-component condensate.
At $g_{12}=g$ the operator $\hat{A}$ of Eq.~(\ref{eq:split}) reduces to $\hat{B}$, whose $n$th orbital is $\phi_{n}$.
The direction $\phi_{n}$ itself, $\psi_{1,2}\propto(1\pm\varepsilon)\phi_{n}$, changes the composition uniformly and would lower the energy as soon as $g_{12}$ exceeds $g$, which is the uniform criterion, but it transfers atoms from one component to the other and is excluded by the orthogonality condition.
The $n$ orbitals below $\phi_{n}$ are the negative directions of the polarization energy, and the lowest polarization that costs energy is along the next orbital, $u_{n+1}^{(0)}$, whose excitation energy $\lambda_{n+1}^{(0)}$ is the spacing between the $n$th and $(n+1)$th orbital energies in the self-consistent potential; both belong to the single-component problem.
At fixed $N_{1}$ and $N_{2}$ the components separate instead by moving in opposite directions in the trap, which costs $\lambda_{n+1}^{(0)}$ and raises the threshold above $g$; this cost is the origin of the trap correction $g_{12}^{*}/g-1$, which is already present for the ground state.

Away from the symmetric point the term $-2\,\delta g\,\phi_{n}^{2}$ lowers the energy of a polarization by the interaction energy that the separation releases; with $\hat{B}$ and the polarization mode held at their values $u_{n+1}^{(0)}$ at the symmetric point, $\lambda_{n+1}$ falls linearly in $\delta g$, with the slope $-2M_{n}$ set by the overlap $M_{n}$, the value of $M$ of Eq.~(\ref{eq:a}) for $u=u_{n+1}^{(0)}$, and the threshold is reached when this release balances the cost,
\begin{equation}
\delta g^{*}\simeq\delta g_{\rm est}\equiv\frac{\lambda_{n+1}^{(0)}}{2M_{n}},
\qquad
M_{n}=\int\big(u_{n+1}^{(0)}\big)^{2}\phi_{n}^{2}\,\dd x.
\label{eq:criterion}
\end{equation}
Both factors follow from the single-component solution at the symmetric point, and no parameter is adjusted.
Figure~\ref{fig:landau}(b) illustrates this balance at $g=1$; the computed coefficient $\alpha=2\lambda_{n+1}$, shown solid, coincides with the estimate, shown dashed, at $g_{12}=g$ but falls faster, since the estimate holds $\phi_{n}$, $\mu_{n}$ and $u_{n+1}$ at their values at the symmetric point, so that the estimate overestimates the threshold.
The estimate is useful because it factorizes the threshold into an excitation energy and an overlap, which can be followed separately as functions of $n$.

\section{Threshold map and the two regimes}
\label{sec:regimes}

\subsection{Thresholds and the estimate}

\begin{table}[t]
\caption{Demixing threshold $g_{12}^{*}(n)$ in oscillator units.}
\label{tab:thresholds}
\begin{ruledtabular}
\begin{tabular}{ccccccc}
$n$ & $g=0.5$ & $1$ & $2$ & $4$ & $8$ & $16$ \\
\hline
0 & 2.0825 & 2.4693 & 3.2968 & 5.0808 & 8.8645 & 16.6811 \\
1 & 2.4885 & 2.8350 & 3.5930 & 5.2840 & 8.9831 & 16.7445 \\
2 & 2.7925 & 3.1204 & 3.8401 & 5.4663 & 9.0936 & 16.8042 \\
3 & 3.0432 & 3.3602 & 4.0551 & 5.6333 & 9.1986 & 16.8615 \\
4 & 3.2603 & 3.5699 & 4.2472 & 5.7878 & 9.2994 & 16.9171 \\
6 & 3.6292 & 3.9293 & 4.5826 & 6.0680 & 9.4899 & 17.0243 \\
8 & 3.9411 & 4.2351 & 4.8726 & 6.3186 & 9.6683 & 17.1274 \\
10 & 4.2149 & 4.5047 & 5.1308 & 6.5466 & 9.8364 & 17.2270 \\
12 & 4.4610 & 4.7478 & 5.3651 & 6.7569 & 9.9959 & 17.3236 \\
14 & 4.6861 & 4.9703 & 5.5806 & 6.9527 & 10.1477 & 17.4175 \\
16 & 4.8943 & 5.1765 & 5.7811 & 7.1365 & 10.2929 & 17.5090 \\
\end{tabular}
\end{ruledtabular}
\end{table}

\begin{figure}[t]
\includegraphics{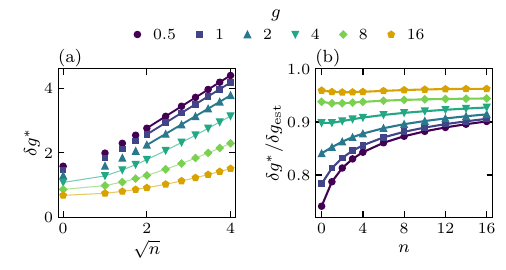}
\caption{Demixing threshold for six values of $g$. (a) Threshold shift $\delta g^{*}=g_{12}^{*}-g$ against $\sqrt{n}$; lines are fits linear in $\sqrt{n}$ over $n=4,6,\dots,16$ at $g\leq2$, and thin lines join the points at $g\geq4$. (b) Ratio of the computed shift to the estimate of Eq.~(\ref{eq:criterion}) against $n$.}
\label{fig:thresholds}
\end{figure}

The threshold rises monotonically with $n$ at every coupling; Table~\ref{tab:thresholds} lists $g_{12}^{*}(n)$ for six values of $g$, and Fig.~\ref{fig:thresholds}(a) plots the shift $\delta g^{*}=g_{12}^{*}-g$ against $\sqrt{n}$, which shows no sign of saturation up to $n=16$.
The estimate of Eq.~(\ref{eq:criterion}) places the threshold too high but captures how it grows with $n$.
Figure~\ref{fig:thresholds}(b) shows the ratio $\delta g^{*}/\delta g_{\rm est}$ of the computed shift to the estimate; at each $g$ it changes by at most $0.16$ between $n=0$ and $n=16$, whereas $\delta g^{*}$ itself grows by a factor between $2.2$ and $2.9$ over the same range.
The ratio lies between $0.74$ and $0.96$, below unity at every $n$ and $g$ computed.
The estimate holds the condensate and the polarization mode at their form at the SU(2) point, whereas the computed $\lambda_{n+1}$ includes their change as $g_{12}$ rises to the threshold, and the solid lines of Fig.~\ref{fig:landau}(b) show that this change lowers it further.
The condensate changes most for the low states at small $g$, where $\delta g^{*}$ exceeds $g$ and $g_{\rm eff}=g+g_{12}$ grows appreciably between the symmetric point and the threshold, and there the ratio is lowest, $0.74$ for $n=0$ at $g=0.5$.
In the TF limit the condensate hardly changes, since $\delta g^{*}\ll g$, but the term $-2\,\delta g\,\phi_{n}^{2}$ of Eq.~(\ref{eq:split}) still deforms the polarization mode, and this deformation alone leaves the ratio for the ground state at about $0.98$.

Since the ratio varies slowly with $n$, the growth of the threshold is carried by the two factors of Eq.~(\ref{eq:criterion}), and which factor carries it depends on how strongly the interaction modifies the trap, which we measure by the mean-field shift of the chemical potential at the symmetric point, $\mu_{n}-(n+\tfrac{1}{2})\hbar\omega$.
At weak intraspecies coupling, where this shift is smaller than $\hbar\omega$, as for every $n\leq16$ at $g\leq1$ and for $n\geq2$ at $g=2$, the self-consistent potential $V+g_{\rm eff}\phi_{n}^{2}$ differs from the trap by a small correction, and the thresholds lie far from the SU(2) point, at $g_{12}^{*}/g$ between $1.9$ and $10$.
In the TF regime, where the shift is much larger than $\hbar\omega$, the interaction flattens the potential inside the cloud, and the healing length is short compared with the size of the cloud, as for $n\leq4$ at $g=16$, where the shift exceeds $5\hbar\omega$, although the notches of the stationary trains are then only four to five healing lengths apart; the couplings $g=4$ and $8$ lie between the two regimes.
For trap frequencies of $2\pi\times30$~Hz and $2\pi\times600$~Hz and a scattering length of $97.7\,a_{0}$, $g=16$ corresponds to $152$ atoms per component, whereas $g\leq2$ means fewer than $20$, so few that exact calculations describe the phase separation of such microscopic mixtures as a crossover~\cite{GarciaMarch2013}, and the mean-field thresholds of this regime are idealizations.

\begin{figure}[t]
\includegraphics{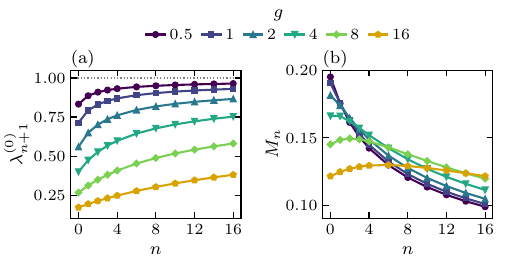}
\caption{The two factors of Eq.~(\ref{eq:criterion}) for six values of $g$. (a) Excitation energy at the SU(2) point; the dotted line marks $\hbar\omega$. (b) Overlap.}
\label{fig:mechanism}
\end{figure}

\subsection{Weak intraspecies coupling}

The intraspecies repulsion is not needed for demixing in this regime; as $g\to0$ the thresholds tend to finite values, from $g_{12}^{*}=1.720$ for $n=0$ to $2.972$ for $n=4$ and $4.626$ for $n=16$, at which the interspecies repulsion alone balances the cost, as kinetic energy alone does in a box without intraspecies interactions~\cite{Wen2012}.
Where $\delta g^{*}$ exceeds $g$, the balance is better evaluated at the threshold than at the SU(2) point.
To first order in the interactions about the linear limit, with $\phi_{n}$ and $u_{n+1}$ the normalized eigenfunctions $h_{n}$ and $h_{n+1}$ of the harmonic oscillator, $h_{n}(x)=(2^{n}n!\sqrt{\pi}\,a_{\rm ho})^{-1/2}H_{n}(x/a_{\rm ho})\,e^{-x^{2}/2a_{\rm ho}^{2}}$ with $H_{n}$ the Hermite polynomials, and $\mu_{n}=(n+\tfrac{1}{2})\hbar\omega+g_{\rm eff}\bar{P}_{n}$, the condition $\lambda_{n+1}=0$ gives
\begin{equation}
g_{12}^{*}\simeq\frac{\hbar\omega+g\,(3\bar{M}_{n}-\bar{P}_{n})}{\bar{M}_{n}+\bar{P}_{n}},
\label{eq:weak}
\end{equation}
with $\bar{M}_{n}=\int h_{n}^{2}h_{n+1}^{2}\,\dd x$ and $\bar{P}_{n}=\int h_{n}^{4}\,\dd x$, in which the interspecies repulsion lowers the energy of the polarization mode by $g_{12}\bar{M}_{n}$ and raises the chemical potential, from which this energy is measured, by $g_{12}\bar{P}_{n}$.

Although the interaction energy at threshold is not small, $g_{12}^{*}\bar{P}_{n}$ lying between $0.6$ and $0.7\,\hbar\omega$ at $g=0$, Eq.~(\ref{eq:weak}) lies between $3\%$ and $6\%$ below the computed thresholds there and within $5\%$ of Table~\ref{tab:thresholds} for $n\geq4$ at $g\leq2$; it lies further below for the lowest states, by $12\%$ for $n=0$ at $g=0.5$ and by up to $29\%$ at $g=2$, where the interaction energy at threshold, $(g+g_{12}^{*})\bar{P}_{n}$, exceeds $\hbar\omega$, and beyond its formal range it serves as an interpolation.
Its denominator, which falls from $0.60$ for $n=0$ to $0.23$ for $n=16$ because $h_{n}^{2}$ and $h_{n+1}^{2}$ spread over a cloud that grows as $\sqrt{n}$ and dilutes, carries the growth of the threshold with $n$.

The same dilution appears in the two factors of Eq.~(\ref{eq:criterion}).
The orbital spacing $\lambda_{n+1}^{(0)}$, which equals $\hbar\omega$ in the trap alone, lies below $\hbar\omega$, since the mean-field potential flattens the trap, and approaches it as the cloud dilutes, as Fig.~\ref{fig:mechanism}(a) shows, whereas the overlap $M_{n}$ falls, as Fig.~\ref{fig:mechanism}(b) shows, its inverse rising by $42\%$ to $44\%$ between $n=4$ and $16$ at $g\leq2$, affinely in the rms radius $R_{n}=(\int x^{2}\phi_{n}^{2}\,\dd x)^{1/2}$ of the cloud.
Since $R_{n}$ grows as $\sqrt{n}$, the computed shifts are well described by $\delta g^{*}\simeq\kappa_{0}+\kappa_{1}\sqrt{n}$ for the levels $n=4,6,\dots,16$ of Table~\ref{tab:thresholds}, with $\kappa_{1}=0.82$, $0.80$ and $0.77$ at $g=0.5$, $1$ and $2$, as the lines in Fig.~\ref{fig:thresholds}(a) show.
This form describes the thresholds over this range of $n$ rather than asymptotically; in the linear limit $M_{n}R_{n}$ grows logarithmically with $n$, and the estimate grows as $\sqrt{n}/\ln n$ far beyond $n=16$.

\subsection{The Thomas--Fermi regime}
\label{sec:tf}

At strong coupling this account fails on both counts, since the orbital spacing falls far below $\hbar\omega$ and rises with $n$, as Fig.~\ref{fig:mechanism}(a) shows, whereas at $g=16$ the overlap hardly changes with $n$, as Fig.~\ref{fig:mechanism}(b) shows, although the cloud doubles in size between $n=2$ and $16$.
The growth of the threshold then comes from the spacing, and its dependence on $n$ changes; the exponent $p$ of a fit $b_{0}+b_{1}n^{p}$ to the computed shifts over the same levels rises from $0.484(4)$ at $g=0.5$ through $0.536(6)$ at $g=2$ and $0.754(8)$ at $g=8$ to $0.856(5)$ at $g=16$, approaching a dependence linear in $n$.
Such a fit describes this range of $n$ rather than the asymptotic dependence.
The spacing at strong coupling is set by the TF profile, in which the interaction flattens the potential inside the cloud and the spacing becomes a spin-wave energy of order $(\hbar\omega)^{2}/\mu$, which falls as $g$ increases.

For the ground state the TF profile reduces the threshold to a universal coefficient, since $\delta g^{*}/g\to0$ and the background keeps its TF form at the symmetric point, with $g_{\rm eff}=2g$ and $\mu=[3g/(2\sqrt{2}\,\hbar\omega a_{\rm ho})]^{2/3}\hbar\omega$ the TF chemical potential at this point; in units of the TF radius the threshold condition $\hat{A}u_{1}=0$ becomes an eigenvalue problem without parameters, whose lowest odd solution, found in Appendix~\ref{app:tf}, gives
\begin{equation}
\frac{g_{12}^{*}(0)}{g}-1\simeq1.822\left(\frac{\hbar\omega}{\mu}\right)^{2}.
\label{eq:tf}
\end{equation}
For $g\geq4$ the computed correction exceeds Eq.~(\ref{eq:tf}) by at most $3\%$, by $2\%$ at $g=16$, and approaches it as $g$ grows.

Expanding the Gaussian variational threshold, Eq.~(20) of Ref.~\cite{Navarro2009}, for small $\hbar\omega/\mu$ gives the same form with the coefficient $25/8$, about $70\%$ above the asymptotic value.
Reference~\cite{Navarro2009} works at fixed chemical potential, and its GP calculations locate the onset of the polarized ground state at $g_{12}/g\simeq1.08$ when the chemical potential of the mixed state at the onset is $5\hbar\omega$; this corresponds to $g=10.0$, where the computed threshold is $g_{12}^{*}/g=1.080$ and Eq.~(\ref{eq:tf}), with the TF chemical potential $4.83\,\hbar\omega$, gives $1.078$.
Since $\mu\propto g^{2/3}$ in 1D, the relative correction $g_{12}^{*}(0)/g-1$ decays as $g^{-4/3}$; it is appreciable only when $\mu$ is a few $\hbar\omega$, and in large condensates the uniform criterion is recovered.

\begin{figure}[t]
\includegraphics{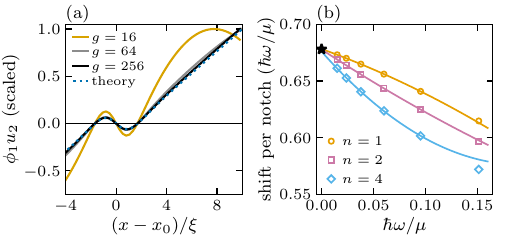}
\caption{Density notches in the TF regime. (a) Spin-density pattern $\phi_{1}u_{2}$ of the threshold mode of the dark--dark soliton against the distance from its notch $x_{0}$ in units of the healing length $\xi$ at the chemical potential of the state, scaled to its largest value within $10\,\xi$ of the notch; the dotted line is the analytical solution near an isolated notch. (b) Relative threshold shift per notch, $[\delta g^{*}(n)/\delta g^{*}(0)-1]/n$ in units of $\hbar\omega/\mu$, against $\hbar\omega/\mu$, with $\mu$ the TF chemical potential of Eq.~(\ref{eq:tf}), for $g=16$ to $512$ (symbols), with quadratic fits over $g=32$ to $512$ (lines); the star marks the coefficient of Eq.~(\ref{eq:notch}).}
\label{fig:domain}
\end{figure}

For the excited states the spacing grows with $n$ because of the dark notches, where the flattened potential keeps narrow wells that bind the $n$ lower orbitals.
On the scale of the cloud an excited state polarizes like the spin dipole of the ground state, and each notch changes the threshold in two ways.
The density vanishes at the notch, and the polarization mode, which cannot use this region, keeps its slope across it and jumps by an amount proportional to this slope; for the mode the notch is thus an excluded segment of effective length $2\sqrt{2}$ healing lengths $\xi=\hbar/(2m\mu)^{1/2}$, which shortens the region available to it and raises its energy; near a central notch the mode reverses sign within $1.7\,\xi$, so that each notch also carries a narrow domain of reversed polarization, as Fig.~\ref{fig:domain}(a) shows.
The notch also expels atoms, which at fixed atom number raises the chemical potential by $\hbar\omega/\sqrt{2}$ and lowers the threshold, since by Eq.~(\ref{eq:tf}) $\delta g^{*}$ falls as $\mu^{-2}$ at fixed $g$.
Matching the threshold mode near a notch to the solution of the TF problem, as Appendix~\ref{app:tf} shows, gives the two contributions of a central notch to the relative change of $\delta g^{*}$ as $+2.092$ and $-1.414$ times $\hbar\omega/\mu$, so that each central notch enlarges $\delta g^{*}$ by the fraction $0.678\,\hbar\omega/\mu$ and, for $n\hbar\omega/\mu\ll1$ and with $\mu$ the TF chemical potential of Eq.~(\ref{eq:tf}),
\begin{equation}
\frac{g_{12}^{*}(n)}{g}-1\simeq1.822\left(\frac{\hbar\omega}{\mu}\right)^{2}\left(1+0.678\,\frac{n\hbar\omega}{\mu}\right).
\label{eq:notch}
\end{equation}

Equation~(\ref{eq:notch}) holds for notches close to the trap center and far apart.
A notch displaced from the center contributes more, since it also removes density where the polarization mode, which vanishes at the center, has a finite amplitude.
The single central notch of the dark--dark soliton meets the condition at every $g$, and its computed fraction, $0.615\,\hbar\omega/\mu$ at $g=16$, lies within $10\%$ of the asymptotic value.
The notches of the stationary trains approach the center only slowly, those of $n=4$ lying within $0.39$ times the TF radius $R_{\rm TF}$ at $g=16$ and within $0.07\,R_{\rm TF}$ at $g=512$, and only four to five healing lengths apart at $g=16$; the computed fractions per notch of all levels nevertheless extrapolate to $0.678\,\hbar\omega/\mu$ as $\mu$ grows, with an empirical leading correction between $-0.25\,n$ and $-0.33\,n$ times $(\hbar\omega/\mu)^{2}$, which makes the step between successive levels fall with $n$, as Fig.~\ref{fig:domain}(b) shows.

Equation~(\ref{eq:notch}) makes the step between successive levels, $\delta g^{*}(n+1)-\delta g^{*}(n)\simeq1.24\,g\,(\hbar\omega/\mu)^{3}$, of third order in $\hbar\omega/\mu$.
Equations~(\ref{eq:tf}) and (\ref{eq:notch}) are asymptotic in $\hbar\omega/\mu$ for the 1D model, which describes an elongated gas as long as $\mu\ll\hbar\omega_{\perp}$; at $g=16$, where $\mu=6.6\,\hbar\omega$, the step formula exceeds the computed steps of Table~\ref{tab:thresholds} by $9\%$ to $24\%$ for $n\leq4$ and by about $50\%$ at $n=16$.
At fixed $n$ the threshold approaches that of the ground state as $\mu$ grows, and since $g$ is proportional to the atom number in a given trap, the steps in $g_{12}/g$, and with them the steps in the interspecies scattering length, fall as the inverse square of the atom number.

\section{The polarized state}
\label{sec:beyond}

\subsection{Supercritical pitchfork}

Symmetry-breaking pitchforks of solitary waves in binary condensates can be supercritical or subcritical~\cite{Katsimiga2017,KatsimigaAppSci2017,Karamatskos2015}, and whether the polarized state grows continuously out of the unpolarized one is decided by the quartic coefficient $\beta$ of $E(\varepsilon)$.
For $\beta>0$ the polarized stationary points, $\varepsilon^{2}=-\alpha/2\beta$, exist where $\alpha<0$ and grow continuously from zero, in a supercritical pitchfork, whereas for $\beta<0$ they fold back to the side $\alpha>0$ and coexist with the unpolarized state, in a subcritical one.

Because the relaxation of the total density and of $\mu_{n}$ contributes at order $\varepsilon^{4}$, $\beta$ does not follow from the path of Eq.~(\ref{eq:ansatz}), and we obtain it instead from the polarized branch itself.
Since the stationary equations are invariant under the exchange of the components combined with $x\to-x$, and $\phi_{n}$ and $u_{n+1}$ have opposite parities, the branch lies in the invariant subspace $\psi_{2}(x)=(-1)^{n}\psi_{1}(-x)$, and we continue it in its projection $\eta=\int u_{n+1}\psi_{1}\,\dd x$ on the threshold mode, which equals $\varepsilon$ to leading order, as Appendix~\ref{app:numerics} describes.

\begin{table}[t]
\caption{Quartic coefficient $\beta$ of $E(\varepsilon)$ at threshold, obtained from the polarized branch, and slope $\lambda_{n+1}'$ of the constrained eigenvalue at threshold, in oscillator units.}
\label{tab:b}
\begin{ruledtabular}
\begin{tabular}{ccccc}
$g$ & $n=0$ & $n=1$ & $n=2$ & $n=4$ \\
\hline
 & \multicolumn{4}{c}{$\beta$} \\
0.5 & 1.816 & 1.889 & 1.905 & 1.905 \\
1   & 1.713 & 1.849 & 1.897 & 1.926 \\
4   & 1.244 & 1.467 & 1.621 & 1.810 \\
16  & 0.628 & 0.712 & 0.792 & 0.939 \\
 & \multicolumn{4}{c}{$-\lambda_{n+1}'$} \\
0.5 & 0.522 & 0.436 & 0.385 & 0.325 \\
1   & 0.488 & 0.426 & 0.383 & 0.327 \\
4   & 0.380 & 0.374 & 0.360 & 0.331 \\
16  & 0.260 & 0.267 & 0.272 & 0.276 \\
\end{tabular}
\end{ruledtabular}
\end{table}

\begin{figure}[t]
\includegraphics{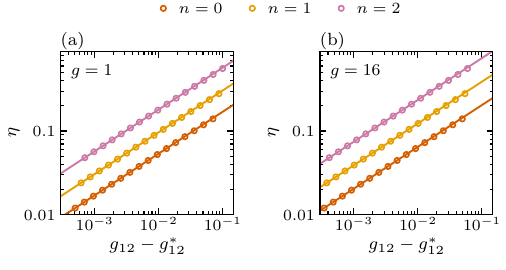}
\caption{Supercritical pitchfork. Projection amplitude $\eta$ of the polarized branch against the distance from threshold at (a) $g=1$ and (b) $g=16$. Symbols are computed branches; lines are Eq.~(\ref{eq:tangent}) with $\beta$ and $\lambda_{n+1}'$ from Table~\ref{tab:b}; for clarity $\eta$ is multiplied by $2$ for $n=1$ and by $4$ for $n=2$.}
\label{fig:pitchfork}
\end{figure}

Table~\ref{tab:b} lists $\beta$ at sixteen points; it is positive at every point, and the pitchfork is supercritical in every case computed.
Near threshold $\lambda_{n+1}\simeq\lambda_{n+1}'(g_{12}-g_{12}^{*})$, where $\lambda_{n+1}'<0$ is the derivative with respect to $g_{12}$, and minimizing $E_{0}+2\lambda_{n+1}\eta^{2}+\beta\eta^{4}$ gives the leading-order laws
\begin{equation}
\begin{aligned}
\eta^{2}&\simeq\frac{|\lambda_{n+1}'|}{\beta}(g_{12}-g_{12}^{*}),\\
E_{\rm unpol}-E_{\rm pol}&\simeq\frac{(\lambda_{n+1}')^{2}}{\beta}(g_{12}-g_{12}^{*})^{2}.
\end{aligned}
\label{eq:tangent}
\end{equation}
The energies of the two branches thus meet at threshold with equal value and equal slope, and the transition is continuous.
Figure~\ref{fig:pitchfork} compares Eq.~(\ref{eq:tangent}) with the branches continued in $\eta$ for $n=0$, $1$ and $2$ at $g=1$ and $16$; the amplitude grows with the mean-field exponent $1/2$ over more than two decades in $g_{12}-g_{12}^{*}$, and the energy difference follows Eq.~(\ref{eq:tangent}) to within $0.5\%$.
Along all sixteen branches $g_{12}$ increases monotonically with $\eta$ up to at least $2g_{12}^{*}$, without folding back at larger amplitude.
The excited states thus demix through the same supercritical pitchfork as the ground state~\cite{Navarro2009}.

\begin{figure*}[t]
\includegraphics{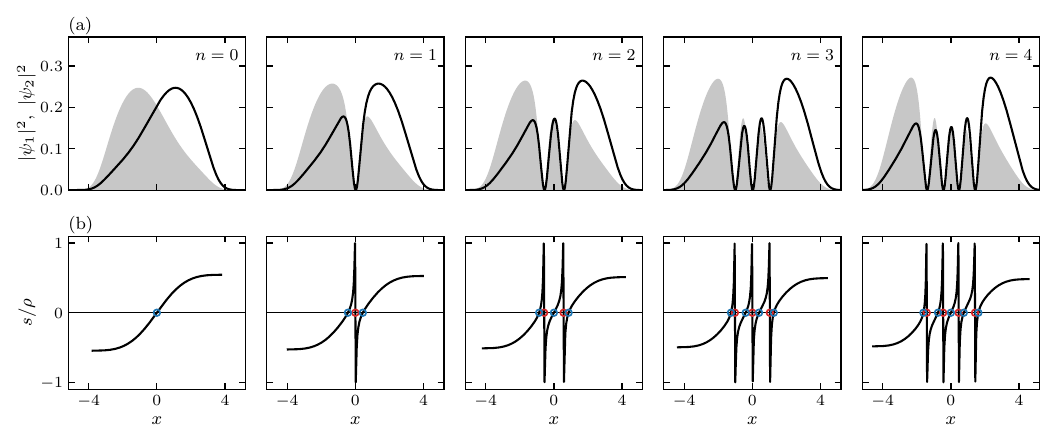}
\caption{Polarized states at $g_{12}=g_{12}^{*}+0.1$ and $g=16$, for $n=0,\dots,4$ from left to right. (a) Densities $|\psi_{1}|^{2}$ (line) and $|\psi_{2}|^{2}$ (shaded). (b) Local polarization $s/\rho$ inside the cloud; circles mark its $2n+1$ sign changes, the $n$ walls at the density notches in red and the $n+1$ walls at the nodes of $u_{n+1}$ in blue.}
\label{fig:order}
\end{figure*}

\subsection{Domain walls}

Near threshold the polarized state divides the condensate into domains of opposite polarization, since its spin density $s\simeq4\varepsilon\phi_{n}u_{n+1}$ vanishes at the $n$ nodes of $\phi_{n}$ and the $n+1$ nodes of $u_{n+1}$.
At threshold $u_{n+1}$ and $\phi_{n}$ are zero modes of $\hat{A}$ and $\hat{B}$, the former by the parity argument of Appendix~\ref{app:constr}, and since $\hat{A}=\hat{B}-2\,\delta g\,\phi_{n}^{2}$ lies below $\hat{B}$ wherever $\phi_{n}\neq0$, the Sturm comparison theorem~\cite{Hartman2002} places at least one node of $u_{n+1}$ strictly inside each of the $n+1$ intervals into which the nodes of $\phi_{n}$ divide the axis, including the two outer intervals, on which both functions decay at infinity, so that their Wronskian vanishes there and the Wronskian proof of the theorem~\cite{Simon2005} applies unchanged, and exactly one, since $u_{n+1}$ has $n+1$ nodes.
The polarization therefore changes sign $2n+1$ times, as Fig.~\ref{fig:order} shows for $n=0$ to $4$, and each sign change is a domain wall.

The two sets of walls differ in character, and Fig.~\ref{fig:order}(b) marks them in different colors; at a node of $u_{n+1}$, blue in the figure, the polarization crosses zero smoothly, as the perturbation does, whereas near a node $x_{0}$ of $\phi_{n}$, red in the figure, the components are $\psi_{1,2}\simeq\phi_{n}'(x_{0})(x-x_{0})\pm\varepsilon u_{n+1}(x_{0})$, which vanish at $x_{0}\mp\varepsilon u_{n+1}(x_{0})/\phi_{n}'(x_{0})$.
Each component thus keeps its dark notch, displaced opposite to that of the other, and between the two notches $s/\rho$ runs from $-1$ to $+1$; this wall is sharp, with a width proportional to $\varepsilon$, and locally the pair of displaced notches is a polarized dark soliton of the kind found in uniform media~\cite{HaeltermanSheppard1994,SheppardKivshar1997}.
In the TF regime the polarized excited state is thus a spin dipole of the whole cloud with, at each notch, a polarized dark soliton flanked by a narrow domain of reversed polarization that ends at the adjacent node of $u_{n+1}$, on the outer side of an off-center notch and on both sides of a central one.
The count persists far from threshold; along the branches continued up to $2g_{12}^{*}$ for $n\leq2$ at $g=1$ and $16$, the polarization changes sign $2n+1$ times at every amplitude.

\section{Dynamical stability}
\label{sec:dynamics}

\begin{figure}[t]
\includegraphics{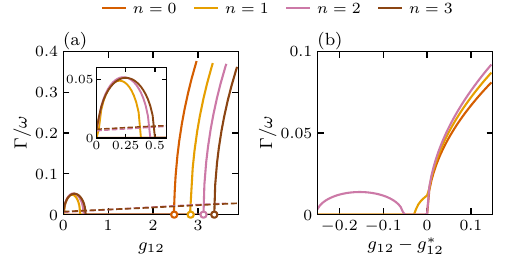}
\caption{Bogoliubov growth rates of the unpolarized states. (a) Spin sector (solid) and charge sector (dashed, $n=2$ and $3$, nearly coincident; stable for $n=0$ and $1$) at $g=1$; circles mark the thresholds, and the inset magnifies the windows at small $g_{12}$. (b) Spin sector at $g=16$ for $n=0$, $1$ and $2$ against the distance from threshold.}
\label{fig:bdg}
\end{figure}

\subsection{Onset of the instability}

Because the unpolarized excited states are saddles of the energy at every coupling, the energy balance of Sec.~\ref{sec:landau} does not decide by itself whether a polarization grows in time, just as a dark soliton, whose energy decreases as it moves and becomes shallower, still oscillates stably in a trap.
Writing the perturbations of the two components as $\delta\psi_{j}=v_{j}+iw_{j}$, the linearized dynamics separates into a charge sector, $\delta\psi_{1}=\delta\psi_{2}$, which carries no polarization, and a spin sector, $\delta\psi_{1}=-\delta\psi_{2}$, which carries it.
In each sector the linearized GP equations read $\hbar\partial_{t}v=\hat{B}w$ and $\hbar\partial_{t}w=-\hat{A}_{\rm charge/spin}v$, so that the real parts are acted on by $\hat{A}_{\rm charge}$ or $\hat{A}_{\rm spin}$ and the imaginary parts by $\hat{B}$, with frequencies
\begin{equation}
\hbar^{2}\Omega^{2}_{\rm charge/spin}=\mathrm{eig}\big(\hat{B}\,\hat{A}_{\rm charge/spin}\big),
\label{eq:bdg}
\end{equation}
where $\mathrm{eig}$ denotes the eigenvalues of the operator product, $\hat{A}_{\rm spin}=\hat{A}$ and $\hat{A}_{\rm charge}=\hat{B}+2g_{\rm eff}\phi_{n}^{2}$.
A frequency with a nonzero imaginary part signals an instability, whose perturbation grows as $e^{\Gamma t}$ with the growth rate $\Gamma=|\mathrm{Im}\,\Omega|$.

For linearizations of this form, the numbers of negative eigenvalues of the two operators fix a weighted count of the unstable modes and of the oscillation modes of negative energy in each sector~\cite{KKS2004,Pelinovsky2005},
\begin{equation}
k_{r}+2k_{c}+2k_{i}=\mathcal{N}_{-}(\hat{A})+\mathcal{N}_{-}(\hat{B}),
\label{eq:krein}
\end{equation}
where $k_{r}$, $k_{c}$ and $k_{i}$ count the positive real eigenvalues of the linearization, its complex quartets and its oscillation modes of negative Krein signature, which lower the energy when excited, and both counts are taken on the complement of $\phi_{n}$, which removes the phase zero mode of $\hat{B}$ and accounts for the conservation of $N_{1}-N_{2}$, as in the constrained count of Refs.~\cite{KKS2004,Pelinovsky2005}.
The count applies where $\hat{A}$ has no zero eigenvalue on this complement and $\mathcal{Q}$ of Appendix~\ref{app:constr} is finite and nonzero, that is, for $g_{12}\neq g$ and $g_{12}\neq g_{12}^{*}$.
In the spin sector $\mathcal{N}_{-}(\hat{B})=n$, since $\phi_{n}$ is the zero mode of $\hat{B}$ and has $n$ nodes, and $\mathcal{N}_{-}(\hat{A})$ equals $n$ below threshold and $n+1$ above it.
The right-hand side of Eq.~(\ref{eq:krein}) therefore changes from $2n$ to $2n+1$ at $g_{12}^{*}$, and since an odd count requires $k_{r}\geq1$, the energetic threshold is also the onset of a non-oscillatory dynamical instability.
Below threshold the count is even, and since the computed spectra contain no real unstable eigenvalue there, it is carried by the $n$ anomalous modes, $k_{i}=n$, the out-of-phase oscillations of the dark notches of the two components, unless an anomalous mode collides in frequency with a mode of positive signature and the pair forms a complex quartet, an oscillatory instability that arises in a Hamiltonian--Hopf bifurcation.
At the SU(2) point $\hat{A}=\hat{B}$ and $\hbar^{2}\Omega^{2}=\mathrm{eig}(\hat{B}^{2})\geq0$, and the spin sector is stable there for every $n$.

\begin{figure}[t]
\includegraphics{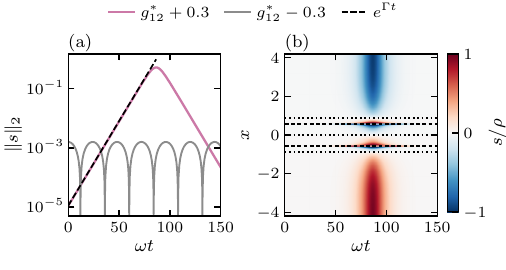}
\caption{Real-time evolution of the $n=2$ state at $g=16$. (a) Polarization amplitude $\|s\|_{2}$ against time above threshold, starting from $1.1\times10^{-5}$, and below threshold, starting from $1.6\times10^{-3}$; the dashed line grows at the linearized rate $\Gamma$. (b) Local polarization inside the cloud against position and time above threshold, $g_{12}=g_{12}^{*}+0.3$; dashed and dotted lines mark the nodes of $\phi_{2}$ and $u_{3}$.}
\label{fig:evolution}
\end{figure}

\subsection{Growth rates}

Figure~\ref{fig:bdg}(a) shows that at $g=1$ the spin-sector growth rate $\Gamma$ becomes nonzero at threshold for every $n$ and rises as $(g_{12}-g_{12}^{*})^{1/2}$, reaching about $0.37\,\omega$ at $g_{12}^{*}+0.5$.
For $n\geq1$ the spin sector carries in addition Hamiltonian--Hopf instabilities near $\Omega\simeq\omega$ at small $g_{12}$, with rates up to $0.05\,\omega$, which for $n=1$ arise from the resonance between the out-of-phase oscillation of the notches and the lowest spin mode~\cite{Morera2018} and lie close to the interval $0.019<g_{12}/g<0.393$ found near the linear limit~\cite{Xu2017}; these windows close at $g_{12}<0.5$ for $n\leq3$, and the spin sector is stable from there up to threshold.
The charge sector, shown dashed, carries for $n\geq2$ the oscillatory instability of multiple dark solitons in a single condensate~\cite{Theocharis2010}, with rates of order $10^{-2}\omega$ at $g=1$, and is stable for $n=0$ and $1$, since at the degenerate frequency $\Omega=\omega$ of the linear limit the mode of positive signature is the Kohn mode, whose frequency stays at $\omega$ exactly, and no quartet forms.

Near threshold the unstable mode has $v\simeq u_{n+1}$, and since $\hat{B}$ is invertible in the parity sector of $u_{n+1}$, eliminating $w=\hbar\Gamma\hat{B}^{-1}v$ and projecting on $u_{n+1}$ gives
\begin{equation}
\hbar^{2}\Gamma^{2}\simeq\frac{|\lambda_{n+1}'|\,(g_{12}-g_{12}^{*})}{\kappa_{n}},
\qquad
\kappa_{n}=\int u_{n+1}\hat{B}^{-1}u_{n+1}\,\dd x,
\label{eq:gamma}
\end{equation}
where $\hat{B}^{-1}u_{n+1}$ is the pattern of relative phase that drives the polarization, and $1/\kappa_{n}$ equals $\lambda_{n+1}^{(0)}$ at the SU(2) point, where $u_{n+1}$ is an eigenfunction of $\hat{B}$.
In the TF regime both factors are small and the rates are smaller; at $g=16$, Fig.~\ref{fig:bdg}(b) shows $\Gamma\simeq0.21\,(g_{12}-g_{12}^{*})^{1/2}\omega$ for $n=0$, with coefficients $0.23$ and $0.24$ for $n=1$ and $2$, as Eq.~(\ref{eq:gamma}) gives with $\lambda_{n+1}'$ from Table~\ref{tab:b}.

Integrating the time-dependent GP equations that follow from Eq.~(\ref{eq:energy}) for the $n=2$ state at $g=16$ tests these rates; we measure the polarization by $\|s\|_{2}=(\int s^{2}\,\dd x)^{1/2}$, which does not depend on the relative phase of the components.
As Fig.~\ref{fig:evolution}(a) shows, at $g_{12}=g_{12}^{*}+0.30$ a seed along the unstable Bogoliubov eigenvector, together with random noise of the same order, grows exponentially over four decades at the linearized rate $\Gamma=0.13\,\omega$, whereas at $g_{12}=g_{12}^{*}-0.30$, below the weak oscillatory window of this state near the threshold, a seed along $u_{n+1}$ oscillates without exceeding its initial value.
Since the dynamics is conservative, $\|s\|_{2}$ overshoots its stationary value, reaching $0.53$ at $\omega t\simeq87$, and returns toward the unpolarized state; whenever the polarization is appreciable it changes sign $2n+1=5$ times at the positions of the walls of the stationary state, as Fig.~\ref{fig:evolution}(b) shows, and the domains of the polarized state therefore form in the dynamics, as domains of opposite composition do when a binary condensate is driven across the miscibility transition~\cite{Sabbatini2011}.

The Bogoliubov spectrum at $g=16$ also has weak oscillatory windows in the spin sector of $n=1$ and $2$ just below the threshold, as Fig.~\ref{fig:bdg}(b) shows, narrow resonances between higher spin modes with rates below $0.014\,\omega$; for $n=1$ the window extends up to the threshold, so that the non-oscillatory growth of this state is preceded in $g_{12}$ by a weak oscillatory one.
Such resonances recur at isolated values of $g$; sampling $g$ at the even values from $8$ to $26$, and for $n=1$ also at $14.7$ and $15.4$, we find them within $0.3$ below the threshold for $n=1$ at $g=16$ and $24$ and for $n=2$ at $g=16$, $18$ and $26$, so that the coupling $g=16$ used in Sec.~\ref{sec:experiment} lies in such an interval.
Far below the SU(2) point the spin sector carries stronger Hamiltonian--Hopf windows, with rates up to $0.26\,\omega$, which close at least $3.8$ below the threshold.

For the trapped dark--dark soliton at $\mu=10\hbar\omega$, with equal chemical potentials of the two components, Ref.~\cite{Morera2018} reported stable and unstable windows alternating up to a value of $g_{12}/g$ close to unity, closer to unity the larger the chemical potential, from a Bogoliubov spectrum computed for $g_{12}/g\leq1$.
At this fixed chemical potential the profile depends on the couplings only through $g+g_{12}$, and the spin sector of the $n=1$ state has its last oscillatory window between $g_{12}/g=0.89$ and $0.93$, is stable above it up to the threshold at $g_{12}^{*}/g=1.023$, and grows exponentially beyond it, which extends that spectrum through the SU(2) point and ends its stable region at the polarization threshold.
At fixed atom number, with $\mu=10\hbar\omega$ reached at the threshold, the threshold is the same and the last window lies between $0.95$ and $0.96$; Eqs.~(\ref{eq:tf}) and (\ref{eq:notch}) give this threshold with $\mu=9.2\,\hbar\omega$, the TF chemical potential of Eq.~(\ref{eq:tf}) for this $g$.

The polarized states themselves are linearly stable near threshold for $n=0$ at both couplings, for $n=1$ at $g=1$ and for $n=2$ at $g=16$, whereas the polarized $n=2$ state at $g=1$ and $n=1$ state at $g=16$ inherit the weak oscillatory instabilities that the unpolarized states carry at threshold, the charge-sector one and the spin-sector window, with rates of $0.025\,\omega$ and $0.012\,\omega$ that vanish by $1.2\,g_{12}^{*}$ and $1.004\,g_{12}^{*}$.
Further from threshold the polarized ground state remains stable up to $2g_{12}^{*}$, whereas the states with $n=1$ and $2$ cross oscillatory windows with rates up to $0.2\,\omega$.

\section{Unequal couplings and population imbalance}
\label{sec:asymmetry}

In real mixtures the intraspecies couplings differ slightly, $g_{11}=g+\zeta$ and $g_{22}=g-\zeta$ with $g$ their arithmetic mean, and the populations are not exactly balanced.
For the $^{87}$Rb pair $|1,-1\rangle$, $|2,-2\rangle$, in which dark--dark and dark--antidark solitons have been observed~\cite{Hoefer2011,Katsimiga2020}, $\zeta/g=7.1\times10^{-3}$~\cite{Hoefer2011}, and for the pair $|1,+1\rangle$, $|2,-1\rangle$, $\zeta/g=2.8\times10^{-2}$~\cite{Tojo2010}.
The interaction energy becomes
\begin{equation}
E_{\rm int}=\int\Big\{\frac{1}{4}\Big[(g+g_{12})\rho^{2}+(g-g_{12})s^{2}\Big]+\frac{\zeta}{2}\rho s\Big\}\dd x,
\label{eq:nsdecomp_asym}
\end{equation}
in which the coefficient of $s^{2}$ is set by the \emph{arithmetic} mean $g$, and a new term couples the total density to the polarization and breaks the exchange symmetry of the components explicitly.

At $\zeta\neq0$ a common profile $\psi_{1}=\psi_{2}=\phi$ no longer solves the stationary equations, since their difference requires $2\zeta\phi^{2}$ to equal the difference of the two chemical potentials, a constant, which a trapped profile cannot satisfy.
Continuing the $\zeta=0$ solution in $\zeta$ gives a nearby branch $\psi_{1,2}=\phi_{n}\pm\chi$ with $\chi=O(\zeta)$, on which the component with the larger self-repulsion expands slightly.
Since $\phi_{n}^{2}$ is even, $\chi$ carries the parity of $\phi_{n}$ and the induced polarization is even, whereas the destabilizing mode $u_{n+1}$ has the opposite parity.
The symmetry broken at threshold is therefore parity, which unequal couplings preserve, and the bifurcation survives as a pitchfork, in which the odd polarization that grows at threshold, with its $2n+1$ walls, adds to the even polarization of the deformed branch and sets the sign pattern once it dominates.

\begin{figure}[t]
\includegraphics{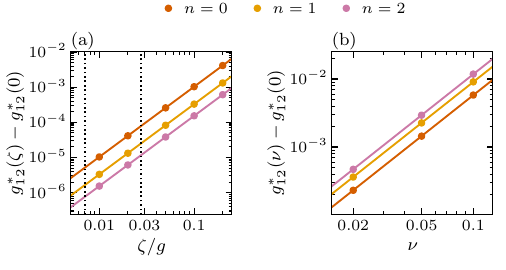}
\caption{Second-order threshold shifts at $g=1$. (a) Unequal intraspecies couplings $g_{11,22}=g\pm\zeta$; dotted lines mark the $^{87}$Rb pairs $|1,-1\rangle$, $|2,-2\rangle$ (left) and $|1,+1\rangle$, $|2,-1\rangle$ (right). (b) Population imbalance $N_{1,2}=N(1\pm\nu)$. Lines are proportional to $\zeta^{2}$ and $\nu^{2}$.}
\label{fig:asymmetry}
\end{figure}

Exchanging the components maps $\zeta\to-\zeta$, and the threshold shift is therefore even in $\zeta$.
Figure~\ref{fig:asymmetry}(a) confirms the quadratic dependence at $g=1$, $g_{12}^{*}(\zeta)-g_{12}^{*}(0)\simeq0.104\,\zeta^{2}$ for $n=0$ in oscillator units, with a coefficient that falls with $n$.
The shift is controlled by the ratio of $\zeta$ to the threshold shift $\delta g^{*}$ rather than by $\zeta/g$, and at $g=16$, where $\delta g^{*}$ is about half its value at $g=1$, its coefficient for $n=0$ is about twice as large.

For a uniform mixture Eq.~(\ref{eq:nsdecomp_asym}) places the demixing at the geometric mean of the intraspecies couplings, which differs from the arithmetic mean at second order, $\sqrt{g_{11}g_{22}}=g-\zeta^{2}/2g+O(\zeta^{4})$, and at that order gives the wrong sign in the trap; substituting it for $g$ in the symmetric threshold predicts that the $n=0$ threshold at $g=1$ and $\zeta=0.20$ falls by $0.016$, whereas the computed threshold rises by $0.004$.
The shift originates instead in the deformation of the unpolarized branch, which neither mean contains.
The quadratic law is an expansion in $\zeta/\delta g^{*}$, which in the TF regime equals $(\zeta/g)(\mu/\hbar\omega)^{2}/1.822$ and reaches unity at $\mu\simeq8\,\hbar\omega$ for the pair $|1,+1\rangle$, $|2,-1\rangle$ and at $\mu\simeq16\,\hbar\omega$ for $|1,-1\rangle$, $|2,-2\rangle$; for the former it reaches about $0.65$ at $g=16$, so that for this pair the shift has to be computed directly; when $\zeta$ exceeds $\delta g^{*}$, as in large condensates, the deformation of the unpolarized branch is no longer small, the TF profiles of the two components acquire different radii, whose edges require a treatment beyond the TF approximation~\cite{Polo2015}, and in the limit of a uniform mixture the geometric-mean criterion is recovered.

A population imbalance $N_{1,2}=N(1\pm\nu)$, with the couplings defined with the mean number $N$ and the components normalized to $1\pm\nu$, enters in the same way, breaking the exchange of the components but not parity, and Fig.~\ref{fig:asymmetry}(b) shows the quadratic dependence of the threshold, with coefficients that exceed those of $\zeta^{2}$ and grow with $n$; an imbalance below $1\%$ still shifts the threshold by less than $1.2\times10^{-4}$ for $n\leq2$.
Since the exchange of the components reverses both $\zeta$ and $\nu$, a term linear in $\nu$ is permitted once $\zeta\neq0$.

\section{Realization with $^{87}$\NoCaseChange{Rb}}
\label{sec:experiment}

The hyperfine pair $|F,m_{F}\rangle=|1,+1\rangle$ and $|2,-1\rangle$ of $^{87}$Rb combines nearly equal intraspecies couplings with a tunable interspecies scattering length.
Its intraspecies scattering lengths are $a_{11}=100.4\,a_{0}$ and $a_{22}=95.00\,a_{0}$, and the interspecies one, $a_{\rm bg}=97.66\,a_{0}$ away from resonance, nearly equals their mean $\bar{a}=97.70\,a_{0}$ and is tuned by a mixed-channel Feshbach resonance near $9.10$~G, near which values of $\Delta a=a_{12}-a_{\rm bg}$ between $-6.5$ and $+8.0\,a_{0}$ have been reported~\cite{Tojo2010}, and an elongated mixture of this pair has been demixed near this resonance~\cite{Nicklas2011}.
Since the dependence of the threshold on $n$ is appreciable only when $\mu$ is a few $\hbar\omega$, we take $\omega=2\pi\times30$~Hz, $\omega_{\perp}=2\pi\times600$~Hz and $N=152$ atoms per component, which gives $g=16$ for the mean scattering length $\bar{a}$.

At $\mu\simeq\hbar\omega_{\perp}/3$ the transverse width of the cloud follows the local density, and we compute the states with the nonpolynomial Schr\"odinger equation (NPSE)~\cite{Salasnich2002}, in which the transverse profile is a Gaussian whose width minimizes the local energy.
The NPSE lowers the chemical potential of the unpolarized ground state at the SU(2) point from $6.64$ to $6.25\,\hbar\omega$, and raises the threshold detunings of the 1D model by $19\%$ and the steps between them by about $20\%$; axisymmetric 3D GP calculations place the thresholds of $n=0$, $1$ and $2$ higher than the NPSE by $0.11\,a_{0}$ and reproduce the steps between them to within $0.002\,a_{0}$.

Table~\ref{tab:exp} lists the thresholds, computed with the unequal couplings and expressed as $\Delta a^{*}=\bar{a}\,g_{12}^{*}/g-a_{\rm bg}$, together with the amount $\Delta a^{*}_{\zeta}$ by which the unequal couplings raise them above the thresholds for equal intraspecies couplings; the ground state demixes at $\Delta a^{*}=5.17\,a_{0}$, and successive levels follow in steps of $0.39(1)\,a_{0}$, within the reported range of $\Delta a$.
The steps are uniform because $\Delta a^{*}_{\zeta}$ falls with $n$, and an imbalance $\nu=\pm0.05$ changes $\Delta a^{*}$ by less than $0.05\,a_{0}$.
The unpolarized state can be prepared from a single-component condensate in its $n$th stationary state, a dark soliton or a train of dark solitons imprinted at their stationary positions~\cite{Becker2008,Theocharis2010}, by a $\pi/2$ pulse at the target field~\cite{Vidanovic2013}, and the polarization can be read out by state-selective imaging; for $n=0$ the centers of mass of the two components separate by $3.4~\mu$m at $2\,a_{0}$ above threshold.
At this atom number quantum fluctuations of the soft mode round the pitchfork; treating $\eta$ and the projection $\chi$ of the relative phase on $u_{n+1}$ as conjugate variables, $[\eta,\chi]=i/4N$, with the phase stiffness $1/\kappa_{n}$ of Eq.~(\ref{eq:gamma}), gives an rms amplitude of $0.064$ at threshold, which the mean-field amplitude reaches $0.07$ to $0.09\,a_{0}$ above threshold, about a fifth of a step.
The unequal couplings polarize the deformed branch itself, with an even pattern that changes sign twice, and the $2n+1$ walls set the sign pattern once the odd polarization dominates, from $0.12$, $0.08$ and $0.06\,a_{0}$ above threshold for $n=0$, $1$ and $2$.

Two features of this resonance make resolving successive levels demanding.
A step corresponds to a few mG of field, larger than the field noise below $1$~mG reached with low-noise current sources~\cite{Zhao2023}, but the scattering lengths determined near the resonance carry a typical accuracy of $\pm1\,a_{0}$~\cite{Tojo2010}, so that the absolute thresholds are known only to a few steps, whereas the differences between levels do not depend on this calibration; at a fixed field the levels can also be compared through their critical atom numbers, which differ by $4\%$ to $6\%$, since $g_{12}/g$ does not depend on $N$ whereas the trap correction does.
Two-body loss near the resonance, with rate coefficients of $1$ to $3\times10^{-13}$~cm$^{3}$/s~\cite{Tojo2010}, removes $8\%$ to $29\%$ of the atoms in $100$~ms, the time over which the polarization grows at the rates of Table~\ref{tab:exp}, and raises the thresholds by a few steps during the hold, so that the separation of the onsets of successive levels has to be compared with calculations that include the measured loss.

The cost of the polarization can also be measured away from the resonance and its loss; at $a_{12}=\bar{a}$ the soft-mode frequency $f_{s}$ is the orbital spacing $\lambda_{n+1}^{(0)}/h$ of Eq.~(\ref{eq:criterion}), and its spectroscopy, as for the spin-dipole mode of a sodium mixture~\cite{Bienaime2016}, would test the growth of this factor with $n$.
The ground state and the dark--dark soliton, whose preparation involves at most one notch at the trap center, are the natural first targets.

\begin{table}[t]
\caption{Predictions for the $^{87}$Rb pair $|1,+1\rangle$, $|2,-1\rangle$ with $N=152$ atoms per component, $\omega=2\pi\times30$~Hz and $\omega_{\perp}=2\pi\times600$~Hz ($g=16$), computed with the NPSE. Threshold detuning $\Delta a^{*}$, its rise $\Delta a^{*}_{\zeta}$ above the value for equal intraspecies couplings, frequency $f_{s}$ of the soft polarization mode at $a_{12}=\bar{a}$, and growth rates $\Gamma_{1}$ and $\Gamma_{2}$ at $1\,a_{0}$ and $2\,a_{0}$ above threshold, the last three computed with equal intraspecies couplings.}
\label{tab:exp}
\begin{ruledtabular}
\begin{tabular}{cccccc}
$n$ & $\Delta a^{*}$ $(a_{0})$ & $\Delta a^{*}_{\zeta}$ $(a_{0})$ & $f_{s}$ (Hz) & $\Gamma_{1}$ (s$^{-1}$) & $\Gamma_{2}$ (s$^{-1}$) \\
\hline
0 & 5.17 & 0.19 & 5.6 & 16 & 23 \\
1 & 5.57 & 0.11 & 6.4 & 17 & 25 \\
2 & 5.96 & 0.07 & 7.0 & 18 & 26 \\
3 & 6.35 & 0.04 & 7.6 & 19 & 27 \\
4 & 6.74 & 0.02 & 8.1 & 20 & 28 \\
\end{tabular}
\end{ruledtabular}
\end{table}

\section{Conclusions}
\label{sec:conclusions}

The unpolarized excited states of a trapped binary condensate demix at an interspecies repulsion that rises with the excitation level, set by the balance between the single-particle cost of separating the components at fixed atom numbers and the interaction energy that the separation releases; evaluated at the SU(2) point, this balance factorizes the threshold into the energy of the polarization mode and its overlap with the condensate.
In the quasi-1D TF limit the trap correction to the ground-state threshold approaches $1.822(\hbar\omega/\mu)^{2}$, and each dark notch near the trap center enlarges it by the fraction $0.678\,\hbar\omega/\mu$ of its value, the net effect of the segment that the notch removes from the polarization mode and of the rise of the chemical potential by the atoms that it expels; at $\mu$ of a few $\hbar\omega$ the computed thresholds give the steps, and these laws their order of magnitude.

The threshold is the onset of an exponential, non-oscillating growth of the polarization, distinct from the oscillatory instabilities that these states can carry, and in every case computed the polarized state emerges through a supercritical pitchfork with $2n+1$ domain walls, $n$ of which sit at the density nodes of the unpolarized state.
Unequal intraspecies couplings and a population imbalance preserve parity and shift the threshold at second order, apart from a term linear in the imbalance once the couplings differ; for the proposed $^{87}$Rb pair the shift from the unequal couplings amounts to half a step for the ground state and makes the steps between levels uniform.
In the hyperfine mixture $|1,+1\rangle$, $|2,-1\rangle$ of $^{87}$Rb the thresholds of the five lowest states lie within the range of interspecies scattering lengths reported near the resonance, in steps of about $0.4\,a_{0}$, and the ground state and the dark--dark soliton are the natural first targets.
In 2D and 3D the nodal points of $\phi_{n}$ become nodal lines and surfaces and the counting that yields $2n+1$ walls changes, whereas the factorization of the threshold into an excitation energy and an overlap is not tied to 1D.

\begin{acknowledgments}
This work was partially supported by the Hebei Natural Science Foundation (A2026502021), the Fundamental Research Funds for the Central Universities (2026MS219), and the Research and Practice Project of Higher Education Teaching Reform of Hebei Province (Nos. 2025GJJG412 and 2026GJJG415).
\end{acknowledgments}

\section*{Data availability}
The data that support the findings of this article are openly available~\cite{Zenodo2026}.

\appendix

\section{Expansion coefficients}
\label{app:a}

Multiplying Eq.~(\ref{eq:single}) by $\phi_{n}$ and integrating gives
\begin{equation}
T_{\phi}+V_{\phi}+g_{\rm eff}P=\mu_{n},
\label{eq:virial}
\end{equation}
with $T_{\phi}=(\hbar^{2}/2m)\int(\phi_{n}')^{2}\,\dd x$, $V_{\phi}=\int V\phi_{n}^{2}\,\dd x$ and $P=\int\phi_{n}^{4}\,\dd x$.
Inserting Eq.~(\ref{eq:ansatz}) into Eq.~(\ref{eq:energy}), the terms linear in $\varepsilon$ cancel between the two components, and with $K_{\phi}=T_{\phi}+V_{\phi}$, $K_{u}=T_{u}+V_{u}$ and $U_{4}=\int u^{4}\,\dd x$ the energy along the path is
\begin{equation}
\begin{aligned}
E(\varepsilon)={}&\frac{2(K_{\phi}+\varepsilon^{2}K_{u})}{1+\varepsilon^{2}}\\
&+\frac{g_{\rm eff}P+(6g-2g_{12})\varepsilon^{2}M+g_{\rm eff}\varepsilon^{4}U_{4}}{(1+\varepsilon^{2})^{2}},
\end{aligned}
\label{eq:pathE}
\end{equation}
exactly.
The intraspecies term contributes $g(P+6\varepsilon^{2}M+\varepsilon^{4}U_{4})$ to the numerator of the second term, since $(\phi_{n}+\varepsilon u)^{4}+(\phi_{n}-\varepsilon u)^{4}=2\phi_{n}^{4}+12\varepsilon^{2}\phi_{n}^{2}u^{2}+2\varepsilon^{4}u^{4}$, and the interspecies term contributes $g_{12}(P-2\varepsilon^{2}M+\varepsilon^{4}U_{4})$, since $(\phi_{n}+\varepsilon u)^{2}(\phi_{n}-\varepsilon u)^{2}=(\phi_{n}^{2}-\varepsilon^{2}u^{2})^{2}$.
Expanding the denominators gives $E_{0}=2K_{\phi}+g_{\rm eff}P$ and
\begin{equation}
\alpha=2(K_{u}-K_{\phi})+(6g-2g_{12})M-2g_{\rm eff}P,
\end{equation}
and eliminating $K_{\phi}+g_{\rm eff}P$ with Eq.~(\ref{eq:virial}) gives Eq.~(\ref{eq:a}).

\section{Constrained eigenvalue problem}
\label{app:constr}

Introducing the multipliers $\lambda$ and $2c$ for the normalization and orthogonality conditions on $u$, we make stationary
\begin{equation}
\mathcal{L}[u]=\int u\hat{A}u\,\dd x-\lambda\Big(\int u^{2}\,\dd x-1\Big)-2c\int\phi_{n}u\,\dd x.
\end{equation}
Under $u\to u+\delta u$, the self-adjointness of $\hat{A}$ gives the first variation $\delta\mathcal{L}=2\int\delta u\,(\hat{A}u-\lambda u-c\,\phi_{n})\,\dd x$, which vanishes for arbitrary $\delta u$ when
\begin{equation}
\hat{A}u=\lambda u+c\,\phi_{n},\qquad \int\phi_{n}u\,\dd x=0.
\label{eq:cep}
\end{equation}
Multiplying Eq.~(\ref{eq:cep}) by $u$ and integrating gives $\lambda=\lambda[u]$, and multiplying it by $\phi_{n}$ gives $c=\int u\,\hat{A}\phi_{n}\,\dd x$.
With the projector $\hat{Q}f=f-\phi_{n}\int\phi_{n}f\,\dd x$, Eq.~(\ref{eq:cep}) is equivalent to $\hat{Q}\hat{A}\hat{Q}u=\lambda u$ with $\hat{Q}u=u$, the eigenvalue problem of $\hat{A}$ restricted to the complement of $\phi_{n}$.
Its eigenfunctions $u_{k}$ are mutually orthogonal and orthogonal to $\phi_{n}$, and since $\hat{A}u_{k}=\lambda_{k}u_{k}+c_{k}\phi_{n}$, one has $\int u_{j}\hat{A}u_{k}\,\dd x=\lambda_{k}\delta_{jk}$, which makes the quadratic energy diagonal, as Sec.~\ref{sec:criterion} uses.

Away from the SU(2) point $\phi_{n}$ is not an eigenfunction of $\hat{A}$, since Eq.~(\ref{eq:single}) gives $\hat{A}\phi_{n}=-2\,\delta g\,\phi_{n}^{3}$, and the constrained spectrum differs from the unconstrained one.
Because $\phi_{n}^{3}$ has the parity of $\phi_{n}$, the multiplier $c=-2\,\delta g\int\phi_{n}^{3}u\,\dd x$ vanishes for every $u$ of the opposite parity, and the threshold mode $u_{n+1}$, which has $n+1$ nodes and this opposite parity, is an eigenfunction of $\hat{A}$ itself at every $g_{12}$, with the same eigenvalue in the constrained and unconstrained spectra.
The unconstrained spectrum contains one additional branch, continuously connected to $\phi_{n}$ and nearly parallel to it, whose eigenvalue, $\simeq-2\,\delta g\int\phi_{n}^{4}\,\dd x$ for small $|\delta g|$, passes through zero at $g_{12}=g$; it is the uniform change of composition that the conservation of $N_{1}$ and $N_{2}$ excludes.
The two counts are related through $\mathcal{Q}=\int\phi_{n}\hat{A}^{-1}\phi_{n}\,\dd x$~\cite{KKS2004,Pelinovsky2005}, since the constrained count $\mathcal{N}_{-}(\hat{A})$ equals the number of negative eigenvalues of $\hat{A}$ on the whole space when $\mathcal{Q}>0$ and falls short of it by one when $\mathcal{Q}<0$.
Below the SU(2) point $\mathcal{Q}>0$ and the two counts coincide, whereas at $g_{12}=g$ the operator $\hat{A}$ annihilates $\phi_{n}$ and $\mathcal{Q}$ passes through a pole, and above this point $\mathcal{Q}<0$ removes the additional branch from the count.
A constrained eigenvalue with the parity of $\phi_{n}$ can cross zero only where $\mathcal{Q}$ does, since the zero mode $u=c\hat{A}^{-1}\phi_{n}$ must then be orthogonal to $\phi_{n}$; the computed count stays $n$ between the SU(2) point and the threshold, so that $\mathcal{Q}$ stays negative there, and at the threshold $\mathcal{Q}$ lies between $-1.01$ and $-0.95$ at $g=1$ and between $-4.3$ and $-3.6$ at $g=16$ for $n=0$, $1$, $2$ and $4$.
The constrained count therefore remains $n$ across the SU(2) point and increases at $g_{12}^{*}$, where an eigenvalue of the opposite parity crosses zero.

\section{Thomas--Fermi limit}
\label{app:tf}

In the TF limit $\phi_{0}^{2}=(\mu-V)/g_{\rm eff}$ inside the cloud, which ends at the TF radius $R_{\rm TF}=(2\mu/m\omega^{2})^{1/2}$.
Writing $u=\phi_{0}y$ and using $\hat{B}\phi_{0}=0$ gives $\hat{B}u=-(\hbar^{2}/2m)(\phi_{0}^{2}y')'/\phi_{0}$, and with $\tilde{x}=x/R_{\rm TF}$ the eigenvalue problem $\hat{B}u=\lambda u$ becomes
\begin{equation}
-\frac{\dd}{\dd\tilde{x}}\Big[(1-\tilde{x}^{2})\frac{\dd y}{\dd\tilde{x}}\Big]=\Lambda(1-\tilde{x}^{2})y,
\qquad
\Lambda=\frac{4\mu\lambda}{(\hbar\omega)^{2}},
\label{eq:tfmode}
\end{equation}
with $y$ bounded at $\tilde{x}=\pm1$.
Its lowest solution, $y=1$ with $\Lambda=0$, is $\phi_{0}$ itself, and the lowest odd solution, with $\Lambda=4.622$, is the spin-dipole mode, whose energy at the symmetric point is $\lambda_{1}^{(0)}=\Lambda(\hbar\omega)^{2}/4\mu$.

Odd modes are eigenfunctions of $\hat{A}$ itself, as Appendix~\ref{app:constr} shows, and for them the threshold condition $\hat{A}u=0$ becomes
\begin{equation}
-\frac{\dd}{\dd\tilde{x}}\Big[(1-\tilde{x}^{2})\frac{\dd y}{\dd\tilde{x}}\Big]=\sigma(1-\tilde{x}^{2})^{2}y,
\qquad
\sigma=\frac{8\mu^{2}\,\delta g}{g_{\rm eff}(\hbar\omega)^{2}}.
\label{eq:tfthr}
\end{equation}
Its lowest odd solution has $\sigma^{*}=7.286$, and with $g_{\rm eff}=2g$ the threshold $\delta g^{*}/g=(\sigma^{*}/4)(\hbar\omega/\mu)^{2}$ gives Eq.~(\ref{eq:tf}).
The estimate of Eq.~(\ref{eq:criterion}) amounts to evaluating the Rayleigh quotient $\int(1-\tilde{x}^{2})(\dd y/\dd\tilde{x})^{2}\,\dd\tilde{x}\big/\!\int(1-\tilde{x}^{2})^{2}y^{2}\,\dd\tilde{x}$ of Eq.~(\ref{eq:tfthr}) with the spin-dipole mode, which gives $7.454$ instead of $\sigma^{*}$ and the coefficient $1.863$.
With $\mu$ the TF chemical potential at $g_{\rm eff}=2g$, the computed coefficient $(g_{12}^{*}(0)/g-1)(\mu/\hbar\omega)^{2}$ exceeds $1.822$ by at most $3\%$ for $g\geq4$, peaks at $g=8$ and approaches it from above for larger $g$, to within $0.2\%$ at $g=256$, with a deviation that falls faster than $\hbar\omega/\mu$ and slower than $(\hbar\omega/\mu)^{2}$ over this range; at smaller $g$, where $\mu$ approaches $\hbar\omega$ and the cloud departs from the TF profile, Eq.~(\ref{eq:tf}) overestimates the correction.

The notches of the excited states add structure on the scale of the healing length $\xi=\hbar/(2m\mu)^{1/2}$.
Near a notch at $x_{\ell}$ the density is $\phi_{n}^{2}\simeq\phi_{0}^{2}(x_{\ell})\tanh^{2}[(x-x_{\ell})/\xi_{\ell}]$, with $\xi_{\ell}=\sqrt{2}\,\xi\,(1-\tilde{x}_{\ell}^{2})^{-1/2}$ the local soliton width, and with $u=\phi_{n}y$ the threshold condition $\hat{A}u=0$ reads $(\phi_{n}^{2}y')'=-(4m\,\delta g/\hbar^{2})\phi_{n}^{4}y$, whose right side is negligible within a few $\xi_{\ell}$ of the notch.
The inner solution is therefore $y=c_{1}(t-\coth t)+c_{2}$, with $t=(x-x_{\ell})/\xi_{\ell}$, which far from the notch behaves as $c_{1}(t\mp1)+c_{2}$ for $t\to\pm\infty$.
On the scale of $R_{\rm TF}$ the outer solution of Eq.~(\ref{eq:tfthr}) thus keeps its slope across the notch and jumps by $[y]=-2\xi_{\ell}y'(x_{\ell})$, the jump produced by excising a segment of length $2\xi_{\ell}$ from the cloud.
For the central notch of an odd state $c_{2}=0$ by parity, the outer solution vanishes at $x=\pm\xi_{\ell}$, and the inner solution changes sign at $t=\coth t$, at $x=1.697\,\xi$, so that between the notch and this node the polarization is reversed; Fig.~\ref{fig:domain}(a) shows that at $g\geq64$ the computed pattern $\phi_{1}u_{2}$ of the dark--dark soliton coincides with the inner solution, and its node stays at $1.70\,\xi$ for $g$ between $16$ and $512$.
For an off-center notch $c_{2}\neq0$, and the node moves toward the notch, to between $0.35$ and $1.35\,\xi$ for $n=2$ and $4$.

Multiplying Eq.~(\ref{eq:tfthr}) by its lowest odd solution $y_{0}$ without notches and integrating by parts on both sides of the notch shows that the jump shifts the eigenvalue by
\begin{equation}
\frac{\delta\sigma}{\sigma^{*}}=\frac{2\xi_{\ell}}{R_{\rm TF}}\,\frac{(1-\tilde{x}_{\ell}^{2})\,y_{0}'(\tilde{x}_{\ell})^{2}}{\int(1-\tilde{x}^{2})\,y_{0}'^{2}\,\dd\tilde{x}}
\label{eq:jump}
\end{equation}
to first order in $\xi_{\ell}/R_{\rm TF}$, where the prime denotes the derivative with respect to $\tilde{x}$.
For a notch off the center the inner region also lacks part of the right side of the threshold condition, since $\phi_{n}^{4}\simeq\phi_{0}^{4}(x_{\ell})\tanh^{4}t$ there, and the missing weight, $\int(1-\tanh^{4}t)\,\dd t=8/3$, adds
\begin{equation}
\frac{\delta\sigma}{\sigma^{*}}=\frac{8\xi_{\ell}}{3R_{\rm TF}}\,\frac{\sigma^{*}(1-\tilde{x}_{\ell}^{2})^{2}\,y_{0}(\tilde{x}_{\ell})^{2}}{\int(1-\tilde{x}^{2})\,y_{0}'^{2}\,\dd\tilde{x}},
\label{eq:weight}
\end{equation}
which vanishes at the center, where $y_{0}=0$; elsewhere the notch removes density where the polarization mode has a finite amplitude, and with it part of the interaction energy that the polarization releases.

For a central notch Eq.~(\ref{eq:jump}) gives $2J\xi_{\ell}/R_{\rm TF}=\sqrt{2}J\hbar\omega/\mu$, with $J=y_{0}'(0)^{2}/\!\int(1-\tilde{x}^{2})y_{0}'^{2}\,\dd\tilde{x}=1.4795$, that is $2.092\,\hbar\omega/\mu$, the result of the Hadamard formula for a Dirichlet boundary moved to $\pm\xi_{\ell}$.
The atoms missing from the notch, $2\xi_{\ell}\phi_{0}^{2}(x_{\ell})$ of the normalized density, raise $\mu$ at fixed atom number by $(1-\tilde{x}_{\ell}^{2})^{1/2}\hbar\omega/\sqrt{2}$, as the computed $\mu_{1}-\mu_{0}=0.71\,\hbar\omega$ at $g=256$ and $512$ confirms for a central notch, and since $\delta g=\sigma g_{\rm eff}(\hbar\omega)^{2}/8\mu^{2}$ at fixed $\sigma$, they lower the threshold by $\sqrt{2}(1-\tilde{x}_{\ell}^{2})^{1/2}\hbar\omega/\mu$.
A central notch therefore contributes $\sqrt{2}(J-1)=0.678$ times $\hbar\omega/\mu$ to the relative threshold shift, the coefficient of Eq.~(\ref{eq:notch}), and a notch at $|\tilde{x}_{\ell}|=0.1$, $0.2$ and $0.3$ contributes $0.76$, $0.98$ and $1.23$ times $\hbar\omega/\mu$, so that a notch displaced from the center raises the threshold.

For the stationary trains $\tilde{x}_{\ell}$ falls only as $(\hbar\omega/\mu)\ln(\mu/\hbar\omega)$, and these position effects, of order $\tilde{x}_{\ell}^{2}$, carry logarithmic factors that are not small at the couplings computed, so that the approach of the computed shifts per notch to $0.678$, shown in Fig.~\ref{fig:domain}(b), rests on extrapolation.
Extrapolated quadratically in $\hbar\omega/\mu$ from $g=32$ to $512$, the computed ratio $[\delta g^{*}(n)/\delta g^{*}(0)-1]\,\mu/n\hbar\omega$ approaches $0.678$ for $n=1$ and $2$ and between $0.676$ and $0.677$ for $n=4$, drifting toward $0.678$ as the fit window moves to larger $g$.
The same fits give the next order, a reduction of the ratio by $0.33$, $0.60$ and $1.0$ times $\hbar\omega/\mu$ for $n=1$, $2$ and $4$, between $0.25\,n$ and $0.33\,n$, which makes the step between successive levels fall with $n$ and brings Eq.~(\ref{eq:notch}) to within $3\%$ of the computed ratios for $n\leq2$ at $g=16$.
Adding to the fits a term $(\hbar\omega/\mu)^{2}\ln(\hbar\omega/\mu)$ changes these coefficients by up to $17\%$, and a term $(\hbar\omega/\mu)\ln(\hbar\omega/\mu)$ instead leaves the limits between $0.677$ and $0.681$; the limit is thus robust, whereas the coefficients of the next order are empirical.
Adjacent notches of the trains lie between $4$ and $5$ healing lengths apart at $g=16$ and between $6$ and $7$ at $g=512$, a spacing that grows only logarithmically with $R_{\rm TF}/\xi$, so that at $g\leq64$ their inner regions overlap and the expansion for separate notches does not apply.
\section{Numerical methods}
\label{app:numerics}

Stationary states are computed on a Fourier spectral grid by a bordered Newton method with damped line search, with the chemical potentials as unknowns alongside the wave functions, and satisfy the stationary equations to $10^{-13}$ in the maximum norm; the box has a length of six times the larger of the linear turning point and the TF radius, and at least $12\,a_{\rm ho}$, the grid spacing is at most a twentieth of the box divided by $n+1$ and a quarter of the healing length, with $160$ to $640$ points, and the boundary densities are below $10^{-20}$.
Distances quoted in healing lengths for computed states use the chemical potential $\mu_{n}$ of each state in place of the TF value.
The constrained spectrum of $\hat{A}$ is obtained by projecting orthogonally to $\phi_{n}$ before diagonalizing, and the threshold is located as the smallest $g_{12}$ at which the number of negative eigenvalues exceeds $n$, by a coarse warm-started sweep followed by bisection, to $10^{-8}$ or better.
For every entry of Table~\ref{tab:thresholds} the count, evaluated at $41$ couplings between $g_{12}=0$ and $0.999\,g_{12}^{*}$ with $\phi_{n}$ continued in $g_{\rm eff}$, equals $n$ at each of them.
For unequal couplings the same counting is applied to the Hessian of the two-component energy on the deformed branch, constrained by both normalizations, whose number of negative eigenvalues rises from $2n$ to $2n+1$ at the threshold, also for the couplings of the $^{87}$Rb pair, where $\zeta/\delta g^{*}\simeq0.65$.

For Appendix~\ref{app:tf} the thresholds at $g=32$ to $512$ are located by a secant iteration on $\lambda_{n+1}(g_{12})$, with the state followed in $g_{\rm eff}$ from $g=16$, on grids with a spacing of a third to a quarter of the healing length, and a grid with half again as many points changes them by less than $10^{-9}$.
For Sec.~\ref{sec:experiment} the same methods are applied to the NPSE, in which the interaction energy density $W=\tfrac{1}{2}(g_{11}|\psi_{1}|^{4}+g_{22}|\psi_{2}|^{4})+g_{12}|\psi_{1}|^{2}|\psi_{2}|^{2}$ of Eq.~(\ref{eq:energy}) is replaced by $\hbar\omega_{\perp}\rho[(1+2W/\hbar\omega_{\perp}\rho)^{1/2}-1]$, and $f_{s}$, $\Gamma_{1}$ and $\Gamma_{2}$ are computed with equal intraspecies couplings.
The 3D check solves the axisymmetric GP equation with $\omega_{\perp}=20\,\omega$ on the half space $x\geq0$, with cell-centred finite differences on $240\times104$ points over $6\,a_{\rm ho}$ axially and $5.8$ transverse oscillator lengths radially, and imposes the parity of $\phi_{n}$ or of the threshold mode through the boundary condition at $x=0$; a grid with half again as many points in each direction changes $\Delta a^{*}$ by less than $0.003\,a_{0}$.

On the invariant subspace $\psi_{2}(x)=(-1)^{n}\psi_{1}(-x)$, on which the two components share one chemical potential, the stationary equation for $\psi_{1}$ is solved together with this chemical potential and $g_{12}$ at fixed projection $\eta$, with $u_{n+1}$ the threshold mode at $g_{12}^{*}$, which keeps the continuation regular at the branch point.
The branch is continued at $16$ amplitudes logarithmically spaced between $0.012$ and $0.14$, and beyond in steps of $0.01$ until $g_{12}>2g_{12}^{*}$, on the grid used for the threshold, which is located there by a secant iteration on $\lambda_{n+1}(g_{12})$.
Fitted over the eight smallest amplitudes, the exponents of $\eta$ and of the energy difference against $g_{12}-g_{12}^{*}$ differ from $1/2$ and $2$ by less than $10^{-3}$ at every point of Table~\ref{tab:b}, by terms of order $\eta^{2}$.
Near threshold $\|\psi_{1}-\psi_{2}\|^{2}\simeq4\eta^{2}$ with $\eta^{2}=-\lambda_{n+1}/\beta$, and the branch lies below the unpolarized state by $\lambda_{n+1}^{2}/\beta$, which gives
\begin{equation}
\beta=-\frac{4\lambda_{n+1}}{\|\psi_{1}-\psi_{2}\|^{2}}=-\frac{\lambda_{n+1}^{2}}{E_{\rm pol}-E_{\rm unpol}},
\label{eq:btrue}
\end{equation}
where $\lambda_{n+1}$ is the constrained eigenvalue of the unpolarized state at the same $g_{12}$.
The two expressions of Eq.~(\ref{eq:btrue}), extrapolated linearly in $\eta^{2}$ over the eight smallest amplitudes, agree to within $2\times10^{-6}$, and a grid with half again as many points changes $\beta$ by less than $2\times10^{-5}$.

The Bogoliubov rates of Eq.~(\ref{eq:bdg}) change by less than $3\times10^{-5}\omega$ when the grid is refined to $512$ points.
For the polarized states the branches are continued in $\eta$ up to $2g_{12}^{*}$, and the linearization about $(\psi_{1},\psi_{2})$ gives $\hbar^{2}\Omega^{2}=\mathrm{eig}(\hat{L}_{-}\hat{L}_{+})$, where $\hat{L}_{-}$ is diagonal in the components, with entries $-(\hbar^{2}/2m)\partial_{x}^{2}+V+g\psi_{j}^{2}+g_{12}\psi_{3-j}^{2}-\mu$, and $\hat{L}_{+}$ adds to it the interaction matrix with entries $2g\psi_{j}^{2}$ and $2g_{12}\psi_{1}\psi_{2}$; rates below $10^{-4}\omega$ come from the two phase zero modes and are discarded.
The real-time integration uses a Strang-split Fourier propagator on $256$ points over $16\,a_{\rm ho}$ with a time step of $0.002/\omega$, which over the full run conserves $N_{1}-N_{2}$ to $2\times10^{-13}$, and the growth rate fitted before saturation agrees with the linearized value to within $10^{-5}$.

\bibliography{main}

\end{document}